# Dispersion laws of the two-dimensional cavity magnetoexciton-polaritons

**S.A.Moskalenko,[a] I.V.Podlesny,[a] E.V.Dumanov,[a,*] M.A.Liberman,[b] B.V.Novikov[c]**
[a]Institute of Applied Physics of the Academy of Sciences of Moldova, Academic Str. 5, Chisinau 2028, Republic of Moldova
[b]Nordic Institute for Theoretical Physics (NORDITA) KTH and Stockholm University, Roslagstullsbacken 23, 106 91 Stockholm , Sweden
[c]St.Petersburg State University, Institute of Physics, Department of Solid State Physics, 1, Ulyanovskaya str., Petrodvorets 198504,St. Petersburg, Russia

**Abstract**. The energy spectrum of the 2D cavity magnetoexciton-polaritons has been investigated previously, using exact solutions for the Landau quantization of conduction electrons and heavy holes provided by the Rashba method [1]. Two lowest Landau quantization levels for electrons and three lowest Landau levels for heavy-holes, lead to the construction of the six lowest magnetoexciton sates. They consist of two dipole-active, two quadrupole-active, and the two forbidden quantum transitions from the ground state of the crystal to the magnetoexciton states. The interaction of the four optical-active magnetoexciton states with the cavity mode photons with a given circular polarization and with well-defined incidence direction leads to the creation of five magnetoexciton-polariton branches. The fifth order dispersion equation is examined by using numerical calculations and the second order dispersion equation is solved analytically, taking into account only one dipole-active magnetoexciton state. The effective polariton mass on the lower polariton branch, the Rabi frequency and the corresponding Hopfield coefficients are determined in dependence on the magnetic field strength, the Rashba spin-orbit coupling parameters and the electron and hole g-factors.

***E-mail**: dum@phys.asm.md

## 1 Introduction

In the previous papers [2, 3] dedicated to the two-dimensional (2D) cavity magnetoexciton-polaritons the dispersion laws of the seven lowest energy branches were described at some discreet values of the external static magnetic and electric fields strengths. This topic is multilateral, because it depends on many parameters and includes four types of interactions such as the Lorentz force, the Rashba spin-orbit coupling, the Coulomb potential and the cavity photon influence. The nonparabolicity of the heavy-hole dispersion law also must be taken into account. In the present paper we continue studies of the energy spectrum in dependence on the broad range of the magnetic and electric field strengths. We consider 2D coplanar electron-hole (e-h) systems in the quantum well (QW) embedded into a microcavity in the presence of the perpendicular strong magnetic and electric fields. The starting phenomena, which determine the main properties of the system, are the Landau quantization (LQ) of the 2D electron and holes states accompanied by the Rashba spin-orbit coupling (RSOC) and by the Zeeman splitting (ZS) effects. In the case of heavy-holes (hhs) there is also the nonparabolicity (NP) of their dispersion low induced simultaneously with the third order chirality terms by the external electric field [4-7]. Exact solutions of the LQ were obtained in the refs [6-8] using the Rashba method [1] and taking into account the RSOC, ZS and PN. In the case of the conduction electrons with the first order chirality terms the exact solutions, supplemented only by adding the influence of the ZS effects, were obtained earlier by Rashba [1]. The fourth order terms in the wave vector components of the type k4 with the amplitude proportional to the electric field strength, as well as the third order chirality terms [4, 5], were introduced in the Refs.[6-8] to avoid the unlimited penetration of the hh energy spectrum inside the semiconductor energy gap. Such penetration is stimulated by the third order chirality terms and it threatens by the collapse of the usual semiconductor band structure, which remains unchanged in reality. The exact solutions mentioned above were used to calculate the Hamiltonians describing the Coulomb electron-electron in the second quantization representation, as well as the electron-radiation interactions. On their base the magnetoexciton energy spectrum and the Hamiltonian describing the magnetoexciton-photon interaction were deduced [2, 9-11]. To calculate the magnetoexciton energy spectra we have taken into account only two lowest Landau levels (LLLs) $(e, R_i)$ with $i = 1, 2$ for electrons and three LLLs $(h, R_j)$ with $j = 1, 2, 3$ for heavy-holes. Their combinations give rise to six lowest magnetoexciton energy levels denoted as $F_n = (e, R_i; h, R_j)$ with $n = 1, 2, ...., 6$. The corresponding energy levels calculated for the GaAs-type QW with their dependences on the magnetic and electric field strengths depicted in Figs.5-7 of Ref. [2]

represent: two dipole-active branches, two quadrupole-active branches, and two forbidden branches in both approximations. These results will be used in the present study. Recently much attention was attracted to the experimental investigations of the influence of a perpendicular magnetic field on the cavity polaritons [12-17]. To the best of our knowledge, the strength of a magnetic field used in experimental studies did not exceed 14 Tesla. From the experimentally observed phenomena one can mention the suppression of the superfluidity, the density dependent Zeeman splitting [14-16], the transverse electric-transverse-magnetic (TE-TM) splitting of the cavity modes, and the Faraday rotation of the plane polarization of the light passing through the microcavity. Authors of [17] used a magnetic field as a tuning parameter for exciton and cavity-photon resonances changing the exciton energy, its oscillator strength, the polariton line-width and the distribution of the polariton population along the dispersion curve. All these phenomena were discussed on the base of the Wannier-Mott exciton structure, taking into account the influence of the magnetic field of arbitrary strength in the frame of the hydrogen interpolation model elaborated in Ref. [18]. The advantage of this model is the possibility to take into account action of the Coulomb potential created by the proton on the electron as well as of the Lorentz force created by the magnetic field. On the contrary, the Landau quantization of electrons and holes is equally taken into account in our approach, whereas the electron-hole binding energy is considered to be much smaller than the cyclotron energies. The most unusual behavior, which is revealed in our model, is related to the nonmonotonous dependences on the magnetic field strength of the energy spectra of the all implicated quasiparticles. They are exclusively dictated by the influence of the perpendicular electric field through the chirality terms determining the RSOC. The cavity photons are the second components of the polaritons. In section three we consider the experimental data of Ref. [17] and compare behavior of the Wannier-Mott excitons and of the magnetoexcitons for different values of the magnetic field.

Unlike to the magnetoexcitons with wave vectors $\vec{k}_{\|}$ oriented in-plane of the QWs, the cavity photons have the wave-vectors $\vec{k}$, which is arbitrarily oriented in the three-dimensional (3D) space of the microcavity and can be expressed in the form $\vec{k} = \vec{a}_3 \vec{k}_z + \vec{k}_{\|}$, where $\vec{a}_3$ is the unit vector perpendicular to the surface of the QW and oriented along the axis of the resonator.

Due to the space quantization the $\vec{k}_z$ component has two discrete values $\vec{k}_z = \pm\pi/L_c$ where $L_c$ is the length of the resonator. Two different values of the wave vector $\vec{k}_{\uparrow\downarrow} = \pm(\pi/L_c)\vec{a} + \vec{a}_{\|}$ determine two directions of the oblique incidence of the cavity photons on the QW, which are embedded in it with the same in plane component $\vec{k}_{\|}$ but with opposite $\vec{k}_z$ components. The gyrotropy effects appearing in the system can be revealed changing the sign $\pm$ of the circular polarizations $\vec{\sigma}^{\pm}_{\vec{k}}$ of the cavity photons at the given orientation of $\vec{k}$, or changing $\vec{k}_{\uparrow}$ by $\vec{k}_{\downarrow}$ at the given sign of the circularity. As it will be shown below, the energy spectrum of 2D cavity polaritons for these conditions is different. According to [17], the magnetic field changes the electron structure of the two-dimensional (2D) Wannier-Mott exciton leading to the shrinkage of its orbital wave function and to the Zeeman effect. As a result, the polariton emission spectra are modified in the same manner. One of them is the shift of the exciton energy level due to the diamagnetic and Zeeman effects, leading to the global shift of the polariton states with the increase of the field strength. The experimental investigations of Ref. [17] were effectuated using the 8 nm-thick $In_xGa_{1-x}As$ single quantum well with $x = 0,04$ characterized by the exciton parameters $E_x = 1,48\text{ev}$, binding energy $7\text{mev}$, and relative motion mass $0,046 m_0$. The detuning $\delta$ between the cavity photon and exciton energies taken at zero magnetic field and at zero in-plane wave vector $k_{\|} = 0$ was chosen close to zero, $\delta = E_c - E_{ex} = -0,09\text{mev}$. For these conditions on the lower polariton (LP) branch the energy shifts are stronger in the range of high values of the in-plane wave vectors $\vec{k}_{\|}$, where the dispersion law is exciton-like, and are smaller in the range of small wave vectors $\vec{k}_{\|}$ where the LP branch is photon-like. The magnetically induced detuning gives rise to a deep dispersion law of the LP branch as it can be seen from fig.1 of Ref.[17]. This detuning leads also to the redistribution of the polariton densities along the dispersion curves. The field induced shrinkage of the exciton wave function directly influences on the exciton-photon interaction energy and changes the exciton oscillator strength, characterizing the quantum transitions from the ground state of the crystal to the Wannier-Mott exciton states, as well as the vacuum Rabi splitting $\Omega$ which in the given sample is equal to $3,5\text{mev}$ $(\Omega = 3,5\text{mev})$. The changes of the exciton wave function influence also the exciton-phonon interaction energy, leading to the variation of the polariton linewidth. So the polariton emission line becomes wider - up to $0,8\text{mev}$ at 14T in the excitonic range of spectrum and up to $0,13\text{mev}$ when the state is mostly photonic. At zero magnetic field the width of polariton line is $0,3\text{mev}$. In the Ref.[17] the Zeeman splitting (ZS) of the LP branch was investigated experimentally. It was observed that the ZS changes with the exciton Hopfield coefficients and can be modeled by independent coupling of the two exciton spin components forming the bright exciton states with the cavity photons. The authors of Ref. [17] have shown the diversity of the magnetoexciton effects already in the linear regime even without nonlinear effects such as the bosonic stimulation and the Bose-Einstein condensation (BEC) of the cavity exciton-polaritons.

The ZS effects for the condition of the BEC of the 2D cavity exciton-polaritons subjected to the influence of the magnetic field up till 9T applied to the microcavity in the Faraday geometry were investigated by Sturm et al. Ref.[19]. The successive polariton condensations have been observed on each of the two polariton spin components with two distinct threshold powers. The nonmonotonous changes of the Zeeman splitting of the polariton branches as well as of the corresponding circular polarized emission lines were measured across the condensation threshold powers. The revealed properties differ essentially from the exciton-spin-Meissner effect predicted in Ref. [14]. The latter is characterized by suppression of the ZS of polariton branches for the magnetic field strength in the range of up till the critical value $B_{cr}$ and is equivalent to the expulsion of the magnetic field from the system in analogy with the physics of superconductors.

The paper is organized as follows. In section 2 the Hamiltonians describing the 2D magnetoexcitons, cavity photons and their interaction are presented Numerical calculations of the dispersion laws and their dependences on the magnetic and electric field strengths are presented in section 3. We conclude in section 4.

## 2 The Hamiltonians of the 2D magnetoexcitons, cavity photons and of their interaction

The Hamiltonians of the 2D magnetoexcitons, of the cavity photons and of their interaction have been obtained in Refs. [2, 3], see e.g. Egs. (69)-(80) of Ref. [2].

$$H_2 = H^0_{mex} + H^0_{ph} + H_{mex-ph}. \tag{1}$$

where $H^0_{mex}$ looks as

$$H^0_{mex} = \sum_{n=1,3,4,5} E_{ex}(\mathrm{F}_n, \vec{\mathrm{k}}_{\|})\hat{\psi}^+_{ex}\left(F_n, \vec{k}_{\|}\right)\hat{\psi}_{ex}\left(F_n, \vec{k}_{\|}\right) \tag{2}$$

It is expressed through the magnetoexciton creation and annihilation operators $\hat{\psi}^+_{ex}\left(F_n, \vec{k}_{\|}\right)$ and $\hat{\psi}_{ex}\left(F_n, \vec{k}_{\|}\right)$ which are determined as follows

$$\hat{\psi}_{ex}^{+}\left(F_n,\vec{Q}\right)=\frac{1}{\sqrt{N}}\sum_t e^{iQ_y t l_0^2} a^{+}_{R_{i,t+\frac{Q_x}{2}}} b^{+}_{R_{j,-t+\frac{Q_x}{2}}} \tag{3}$$

$$F_n=\left(e,R_i;h,R_j\right);i=1,2;\ j=1,2,3;n=1,2,..,6$$

where $a^{+}_{R_{i,t}}$ and $b^{+}_{R_{j,t}}$ are the electron and hole creation operators in the states $\left(e,R_i\right)$ and $\left(h,R_j\right)$ correspondingly, whereas $l_0$ is the magnetic length.

The magnetoexciton energy branches $E_{ex}\left(F_n,\vec{k}_n\right)$ depend on the 2D in-plane wave vector $\vec{k}_{\|}$ and following the formula (66) of Ref. [2] consist from some parts

$$E_{ex}\left(F_n,\vec{k}_{\|}\right)=E_g^0+E_e\left(R_i\right)+E_h\left(R_j\right)-I_{ex}\left(F_n,0\right)+\frac{\hbar^2\vec{k}_{\|}^2}{2M\left(F_n,B\right)} \tag{4}$$

Here $E_g^0$ is the semiconductor energy gap, $E_e\left(R_i\right)$ and $E_n\left(R_j\right)$ are the energy levels of the electron and of the heavy-hole (hh) in their states of Landau quantization accompanied by the RSOC, ZS and NP of the dispersion law. $I_{ex}\left(F_n,0\right)$ is the ionization potential of the magnetoexciton with wave vector $\vec{k}_{\|}=0$, arising due to the Coulomb electron-hole attraction and $M\left(F_n,B\right)$ is the effective magnetic mass. The values $E_{ex}\left(F_n,0\right)$ and $M\left(F_n,\mathrm{B}\right)$ in dependence on the magnetic field strength were represented in the figures 5-7 and 8, correspondingly, of Ref. [2].

The Hamiltonian of cavity photons with circular polarizations $\vec{\sigma}_{\vec{k}}^{\pm}$ has the form

$$H_{ph}^0=\sum_{\vec{k}_{\|},\vec{k}_z=\pm\frac{\pi}{L_c}}\hbar\omega_{\vec{k}}[\left(C_{\vec{k},+}\right)^{+}\mathrm{C}_{\vec{k},+}+\left(C_{\vec{k},-}\right)^{+}\mathrm{C}_{\vec{k},-}] \tag{5}$$

The photon creation and annihilation operators with different circular $\sigma_{\vec{k}}^{\pm}$ and linear $\vec{e}_{\vec{k},1},\vec{e}_{\vec{k},2}$ polarizations are determined as:

$$
\begin{aligned}
&C_{\vec{k},\pm} = \frac{1}{\sqrt{2}}\left(C_{\vec{k},1} \pm iC_{\vec{k},2}\right); \quad \left(C_{\vec{k},\pm}\right)^{+} = \frac{1}{\sqrt{2}}\left(C^{+}_{\vec{k},1} \mp iC^{+}_{\vec{k},2}\right) \\
&\vec{\sigma}^{\pm}_{\vec{k}} = \frac{1}{\sqrt{2}}\left(\vec{e}_{\vec{k},1} \pm \vec{e}_{\vec{k},2}\right); \ \left(\vec{e}_{\vec{k},\mathrm{i}} \cdot \vec{k}\right) = 0; \ i = 1,2; \left(\vec{\sigma}^{\pm}_{\vec{k}} \cdot \vec{k}\right) = 0 \\
&\sum_{i=1}^{2} c_{\vec{k},i}\vec{e}_{\vec{k},i} = C_{\vec{k},-}\vec{\sigma}^{+}_{\vec{k}} + C_{\vec{k},+}\vec{\sigma}^{-}_{\vec{k}} \\
&\sum_{i=1}^{2} C^{+}_{\vec{k},i}\vec{e}_{k,i} = \left(C_{\vec{k},-}\right)^{+}\vec{\sigma}^{-}_{\vec{k}} + \left(C_{\vec{k},+}\right)^{+}\vec{\sigma}^{+}_{\vec{k}} \\
&\vec{\sigma}_{M_h} = \frac{1}{\sqrt{2}}\left(\vec{a}_1 \pm i\vec{a}_2\right); \ \vec{k} = k_z\vec{a}_3 + \vec{k}_{\|}; \ \vec{k}_{\|} = \vec{a}_1 k_x + \vec{a}_2 k_y
\end{aligned}
\tag{6}
$$

The circular polarization vectors $\vec{\sigma}_{M_h}$ characterize the quantum states of the heavy-holes in the frame of the p-type valence band with the magnetic numbers $M_h = \pm 1$. They are expressed through the in-plane unit vectors $\vec{a}_1$ and $\vec{a}_2$. The unit vector $\vec{a}_3$ is perpendicular to the QW plane. The energy spectrum of the cavity photons propagating with arbitrary 3D wave vector $\vec{k} = \vec{a}_3\vec{k}_z + \vec{k}_n$ inside the resonator with the length $L_c$ and with the spacer refractive index $n_c$ is $\hbar\omega_{\vec{k}} = \hbar c\,|\vec{k}|/n_c$.

In the case of a small in-plane component $|\vec{k}|$ of the photon wave vector $\vec{k}$ as compared with the quantized longitudinal component $|\vec{k}| = \pi/L_C$ the cavity photon dispersion law is given by the formula (64) of the Ref. [2]

$$
\begin{aligned}
&\hbar\omega_{\vec{k}} = \frac{\hbar c}{n_c}\sqrt{\left(\frac{\pi}{L_c}\right)^2 + \vec{k}_{\|}^2} \approx \hbar\omega_c + \frac{\hbar^2\vec{k}_{\|}^2}{2m_c} = \hbar\omega_c\left(1 + \frac{x^2}{2}\right), x = \frac{|\vec{k}|\,L_c}{\pi} < 1 \\
&\omega_c = \frac{c\pi}{n_c L_c}, \ m_c = \frac{\hbar\pi n_c}{cL_c}
\end{aligned}
\tag{7}
$$

The magnetoexciton-cavity-photon interaction Hamiltonian taking into only the resonance terms and neglecting by the antiresonance ones is described by the formula (79) of Ref. [2]

$$H_{mex-ph} = \sum_{\vec{k}_{\parallel},\vec{k}_z=\pm\frac{\pi}{L_c}} \Big\{ \varphi\left(F_1,\vec{k}_{\parallel}\right)\left[C_{\vec{k},-}\left(\vec{\sigma}_{\vec{k}}^{+}\cdot\vec{\sigma}_{-1}^{*}\right)+C_{\vec{k},+}\left(\vec{\sigma}_{\vec{k}}^{-}\cdot\vec{\sigma}_{-1}^{*}\right)\right]\hat{\psi}_{ex}^{+}\left(F_1,\vec{k}_{\parallel}\right)+\varphi\left(F_4,\vec{k}_{\parallel}\right)$$
$$\left[C_{\vec{k},-}\left(\vec{\sigma}_{\vec{k}}^{+}\cdot\vec{\sigma}_{1}^{*}\right)+C_{\vec{k},+}\left(\vec{\sigma}_{\vec{k}}^{-}\cdot\vec{\sigma}_{1}^{*}\right)\right]\hat{\psi}_{ex}^{+}\left(F_4,\vec{k}_{\parallel}\right)+ \tag{8}$$
$$+\varphi\left(F_3,\vec{k}_{\parallel}\right)\left[C_{\vec{k},-}\left(\vec{\sigma}_{\vec{k}}^{+}\cdot\vec{\sigma}_{1}^{*}\right)+C_{\vec{k},+}\left(\vec{\sigma}_{\vec{k}}^{-}\cdot\vec{\sigma}_{1}^{*}\right)\right]\psi_{ex}^{+}\left(F_3,\vec{k}_{\parallel}\right)+$$
$$+\varphi\left(F_5,\vec{k}_{\parallel}\right)\left[C_{\vec{k},-}\left(\vec{\sigma}_{\vec{k}}^{+}\cdot\vec{\sigma}_{-1}^{*}\right)+C_{\vec{k},+}\left(\vec{\sigma}_{\vec{k}}\cdot\vec{\sigma}_{-1}^{*}\right)\right]\psi_{ex}^{+}\left(F_5,\vec{k}_{\parallel}\right)+h.c.\Big\}$$

The circular polarization vectors $\vec{\sigma}_{\vec{k}}^{\pm}$ of the light propagating inside the cavity with arbitrary oriented wave vector $\vec{k}$ as regards the QW embedded in it can be expressed in terms of two linear polarization vectors $\vec{e}_{\vec{k},1}=\vec{s}_{\vec{k}}$ and $\vec{e}_{\vec{k},2}=\vec{t}_{\vec{k}}$ following the formula (72) of the Ref.[2] in the way

$$\begin{aligned}
&\vec{\sigma}_{\vec{k}}^{\pm}=\frac{1}{\sqrt{2}}\left(\vec{s}_{\vec{k}}\pm i\vec{t}_{\vec{k}}\right)=\left(\vec{\sigma}_{\vec{k}}^{\mp}\right)^{*}\\
&\vec{s}_{\vec{k}}=\vec{a}_3\frac{\left|\vec{k}_{\parallel}\right|}{\left|\vec{k}\right|}-\frac{\vec{k}_{\parallel}\cdot\vec{k}_z}{\left|\vec{k}\right|\left|\vec{k}_{\parallel}\right|}=\vec{s}_{-\vec{k}}\\
&\vec{t}_{\vec{k}}=\frac{\vec{a}_1\vec{k}_y-\vec{a}_2\vec{k}_x}{\left|\vec{k}_{\parallel}\right|}=-\vec{t}_{-\vec{k}}\\
&\left(\vec{k}\cdot\vec{s}_{\vec{k}}\right)=\left(\vec{k}\cdot\vec{t}_{\vec{k}}\right)=\left(\vec{s}_{\vec{k}}\cdot\vec{t}_{\vec{k}}\right)=\left(\vec{\sigma}_{\vec{k}}^{\pm}\cdot\vec{k}\right)=0\\
&\left|\vec{s}_{\vec{k}}\right|=\left|\vec{t}_{\vec{k}}\right|=\left|\sigma_{\vec{k}}^{\pm}\right|=\left|\vec{\sigma}_{M_h}\right|=1
\end{aligned} \tag{9}$$

They lead to the resultant expression

$$\vec{\sigma}_{\vec{k}}^{\pm}=\frac{1}{\sqrt{2}\left|\vec{k}\right|\left|\vec{k}_{\parallel}\right|}\left\{\vec{a}_3\left|\vec{k}_{\parallel}\right|^2+\vec{a}_1\left(-\vec{k}_x\vec{k}_z\pm i\vec{k}_y\left|\vec{k}\right|\right)+\vec{a}_2\left(-k_yk_z\mp ik_x\left|\vec{k}\right|\right)\right\} \tag{10}$$

and to the geometric selection rules

$$\begin{aligned}
&\left|\left(\vec{\sigma}_{\vec{k}}^{\pm}\cdot\vec{\sigma}_{1}^{*}\right)\right|^2=\frac{1}{4\left|\vec{k}\right|^2}\left(\vec{k}_z\pm\left|\vec{k}\right|\right)^2\\
&\left|\left(\vec{\sigma}_{\vec{k}}^{\pm}\cdot\vec{\sigma}_{-1}^{*}\right)\right|^2=\frac{1}{4\left|\vec{k}\right|^2}\left(\vec{k}_z\mp\left|\vec{k}\right|\right)^2
\end{aligned} \tag{11}$$

They depend not only on the sign $\pm$ of the circular polarization vectors $\vec{\sigma}_{\vec{k}}^{\pm}$, but also on the sign of the longitudinal projection $\vec{k}_z = \pm\pi/L_c$. One can change the geometric selection rules changing the sign of the light circular polarization at a given sign of the $\vec{k}_z$ projection, or changing the sign of the $\vec{k}_z$ projection at a given sign of the circular polarization. Changing simultaneously the both signs, the geometric selection rules remain unchanged. These rules can be better represented introducing the photon wave vectors $\vec{k}_{\uparrow,\downarrow} = \pm(\pi/L_c)\vec{a}_3 + \vec{k}_{\parallel}$ and paying attention, that the geometric selection rules are the same in the cases of the circular polarizations $\vec{\sigma}_{\vec{k}_{\uparrow}}^{\pm}$ and $\vec{\sigma}_{\vec{k}_{\downarrow}}^{\mp}$. It means that the geometric selection rules depend not only on the sign of the circular polarization $\pm$, but also on the direction along which the photon incidents on the surface of the QW, whether along the magnetic field direction or from the opposite side.

In the approximation of the small values $\left|\vec{k}_{\parallel}\right|$ we have

$$\begin{aligned}
&\left|\left(\vec{\sigma}_{\vec{k}_{\uparrow}}^{+}\cdot\vec{\sigma}_{1}^{*}\right)\right|^{2} = \left|\left(\vec{\sigma}_{\vec{k}_{\downarrow}}^{-}\cdot\vec{\sigma}_{1}^{*}\right)\right|^{2} = \left|\left(\vec{\sigma}_{\vec{k}_{\uparrow}}^{-}\cdot\vec{\sigma}_{-1}^{*}\right)\right|^{2} = \left|\left(\vec{\sigma}_{\vec{k}_{\downarrow}}^{+}\cdot\vec{\sigma}_{-1}^{*}\right)\right|^{2} \approx \left(1-\frac{x^{2}}{2}+\frac{7}{16}x^{4}\right);\ |x|<1 \\
&\left|\left(\vec{\sigma}_{\vec{k}_{\uparrow}}^{+}\cdot\vec{\sigma}_{-1}^{*}\right)\right|^{2} = \left|\left(\vec{\sigma}_{\vec{k}_{\downarrow}}^{-}\cdot\vec{\sigma}_{-1}^{*}\right)\right|^{2} = \left|\left(\vec{\sigma}_{\vec{k}_{\uparrow}}^{-}\cdot\vec{\sigma}_{1}^{*}\right)\right|^{2} = \left|\left(\vec{\sigma}_{\vec{k}_{\downarrow}}^{+}\cdot\vec{\sigma}_{1}^{*}\right)\right|^{2} \approx \frac{x^{4}}{16}
\end{aligned} \tag{12}$$

The coefficients $\varphi\left(F_n, \vec{k}_{\parallel}\right)$ determining the Hamiltonian (8) were expressed by the formulas (80) of Ref. [2] as:

$$
\begin{aligned}
&\varphi\left(F_1,\vec{k}_{\parallel}\right)=-\phi_{cv}a_0^{-*}d_0^{-*}\\
&\left|\varphi\left(F_1,\vec{k}_{\parallel}\right)\right|^2=\left|\phi_{cv}\right|^2\left|a_0^-\right|^2\left|d_0^-\right|^2,\\
&\varphi\left(F_4,\vec{k}_{\parallel}\right)=\phi_{cv};\left|\varphi\left(F_4,\vec{k}_{\parallel}\right)\right|^2=\left|\phi_{cv}\right|^2,\\
&\varphi\left(F_3,\vec{k}_{\parallel}\right)=\phi_{cv}b_1^{-*}\left(\frac{-k_x+ik_y}{\sqrt{2}}\right)l_0\\
&\left|\varphi\left(F_3,\vec{k}_{\parallel}\right)\right|^2=\left|\phi_{cv}\right|^2\left|b_1^-\right|^2\frac{\left|\vec{k}_{\parallel}\right|^2 l_0^2}{2}\\
&\varphi\left(F_5,\vec{k}_{\parallel}\right)=-\phi_{cv}a_0^{-*}d_1^{-*}\left(\frac{k_x+ik_y}{\sqrt{2}}\right)l_0\\
&\left|\varphi\left(F_5,\vec{k}_{\parallel}\right)\right|^2=\left|\phi_{cv}\right|^2\left|a_0^-\right|^2\left|d_1^-\right|^2\frac{\left|\vec{k}_{\parallel}\right|^2 l_0^2}{2}\\
&\phi_{cv}=\frac{e}{m_0 l_0}\sqrt{\frac{\hbar n_c}{\pi c}}P_{cv(0)}\\
&\left|\phi_{cv}\right|^2=\frac{\hbar n_c}{\pi c}\left(\frac{eP_{cv(0)}}{m_0 l_0}\right)^2\\
&f_{osc}=\left|\frac{\phi_{cv}}{\hbar\omega_c}\right|^2=\frac{\hbar n_c}{\pi c}\left|\frac{eP_{cv(0)}}{m_0 l_0\hbar\omega_c}\right|^2
\end{aligned}
\tag{13}
$$

One can observe that the probabilities of the dipole-active quantum transitions in the states $F_1$ and $F_4$ determined by the expression $\left|\varphi\left(F_1,\vec{k}_{\parallel}\right)\right|^2$ and $\left|\varphi\left(F_4,\vec{k}_{\parallel}\right)\right|^2$ are proportional to $\left|\phi_{cv}\right|^2=l_0^{-2}\approx B$ and have a linear dependence on the magnetic field strength B. The magnetoexciton states $F_3$ and $F_5$ are quadrupole-active and the corresponding probabilities $\left|\varphi\left(F_3,\vec{k}_{\parallel}\right)\right|^2$ and $\left|\varphi\left(F_5,\vec{k}_{\parallel}\right)\right|^2$ are proportional to $\left|\phi_{cv}\right|^2 l_0^2\left|\vec{k}_{\parallel}\right|^2$. They are proportional to the square modulus of the in-plane wave vector $\left|\vec{k}_{\parallel}\right|^2$, but do not depend on the magnetic field strength B.

The equations of motions for four magnetoexciton annihilation operators $\hat{\psi}_{ex}\left(F_n,\vec{k}_{\|}\right)$ with $n=1,3,4,5$ as well as for the cavity-photon annihilation operator $C_{\vec{k},\pm}$ were deduced in Ref. [2], see Eqs. (81) and (82). For the stationary conditions they look as

$$\begin{aligned} &i\hbar\frac{d}{dt}\hat{\psi}_{ex}\left(F_n,\vec{k}_{\|}\right)=\left[\hat{\psi}_{ex}\left(F_n,\vec{k}_{\|}\right),\hat{H}_2\right]=\hbar\omega\hat{\psi}_{ex}\left(F_n,\vec{k}_{\|}\right),\\ &i\hbar\frac{d}{dt}C_{\vec{k},\pm}=\left[C_{\vec{k},\pm}\right]=\left[C_{\vec{k},\pm},\hat{H}_2\right]=\hbar\omega C_{\vec{k},\pm} \end{aligned} \tag{14}$$

If there are only photons with one circular polarization either $\vec{\sigma}_{\vec{k}}^{+}$ or $\vec{\sigma}_{\vec{k}}^{-}$ in the microcavity, and the Hamiltonians (5) and (8) contain only the corresponding photon operators, the equations of motion lead to the following relations between the exciton and the photon operators

$$\begin{aligned} &\hat{\psi}_{ex}\left(F_i,\vec{k}_{\|}\right)=C_{\vec{k},\pm}\frac{\left(\vec{\sigma}_{\vec{k}}^{\mp}\cdot\vec{\sigma}_{-1}^{*}\right)\varphi\left(F_i,\vec{k}_{\|}\right)}{\hbar\omega-E_{ex}\left(F_i,\vec{k}_{\|}\right)};\ i=1,5,\\ &\hat{\psi}_{ex}\left(F_j,\vec{k}_{\|}\right)=C_{\vec{k},\pm}\frac{\left(\vec{\sigma}_{\vec{k}}^{\mp}\cdot\vec{\sigma}_{1}^{*}\right)\varphi\left(F_j,\vec{k}_{\|}\right)}{\hbar\omega-E_{ex}\left(F_j,\vec{k}_{\|}\right)};\ \mathrm{j}=3,4 \end{aligned} \tag{15}$$

and to the dispersion equations (see Eq.(84) of Ref. [2]):

$$\begin{aligned} \left(\hbar\omega-\hbar\omega_{\vec{k}}\right)=&\frac{\left|\left(\vec{\sigma}_{\vec{k}}^{\mp}\cdot\vec{\sigma}_{-1}^{*}\right)\right|\left|\varphi\left(F_1,\vec{k}_{\|}\right)\right|^2}{\hbar\omega-E_{ex}\left(F_1,\vec{k}_{\|}\right)}+\frac{\left(\vec{\sigma}_{\vec{k}}^{\mp}\cdot\vec{\sigma}_{1}^{*}\right)\left|\varphi\left(F_4,\vec{k}_{\|}\right)\right|^2}{\hbar\omega-E_{ex}\left(F_4,\vec{k}_{\|}\right)}+\\ &+\frac{\left|\left(\vec{\sigma}_{\vec{k}}^{\mp}\cdot\vec{\sigma}_{1}^{*}\right)\right|^2\left|\varphi\left(F_3,\vec{k}_{\|}\right)\right|^2}{\hbar\omega-E_{ex}\left(F_3,\vec{k}_{\|}\right)}+\frac{\left|\vec{\sigma}_{\vec{k}}^{\mp}\cdot\vec{\sigma}_{-1}^{*}\right|^2\left|\varphi\left(F_5,\vec{k}_{\|}\right)\right|^2}{\hbar\omega-E_{ex}\left(F_5,\vec{k}_{\|}\right)} \end{aligned} \tag{16}$$

These five order algebraic dispersion equations describe the interaction of four magnetoexciton branches with the cavity photon branch with a given circular polarization either $\vec{\sigma}_{\vec{k}}^{-}$ or $\vec{\sigma}_{\vec{k}}^{+}$.

The energy of the cavity mode $\hbar\omega_c$ in Refs.[2, 3] was tuned to the energy of the selected magnetoexciton energy level at the given value of the magnetic field strength $B$ to investigate the polariton dispersion law in the vicinity of this level. Two different values of the magnetic field strength 20 and 40 Tesla were used and two different values of the magnetoexciton energy levels were considered. In these calculations the different cavity mode energies $\hbar\omega_c\left(B\right)$ were taken to obtain the same relative detuning between the magnetoexciton and the cavity mode energies.

At the same time, the case with a fixed value of the cavity mode energy $\hbar\omega_c$ independent on the magnetic In Refs.[2, 3] field strength is interesting. In this case the detuning between two mentioned energies will depend on $B$.

To this end, the detuning of the cavity mode energy as regard the semiconductor band energy gap $E_g^0$ is introduced and the following denotations are used

$$\begin{aligned}
&\hbar\omega = \hbar\omega_c + E;\ \frac{\hbar\omega}{\hbar\omega_c} = 1+\varepsilon;\ \varepsilon = \frac{E}{\hbar\omega_c}\\
&\Delta = \hbar\omega_c - E_g^0;\ \delta = \frac{\Delta}{\hbar\omega_c};\ \frac{E_g^0}{\hbar\omega_c} = 1-\delta;\ x = \frac{\left|\vec{k}_{\|}\right| L_c}{\pi},\\
&\tilde{E}_{ex}\left(F_n,\vec{k}_{\|}\right) = E_{ex}\left(F_n,\vec{k}_{\|}\right) - E_g^0 = \tilde{E}_{ex}\left(F_n,0\right) + \frac{\hbar^2\vec{k}_{\|}^2}{2M\left(F_n,B\right)},\\
&\tilde{E}_{ex}\left(F_n,0\right) = E_{ex}\left(F_n,0\right) - E_g^0,\\
&\frac{\hbar\omega - E_{ex}\left(F_n,\vec{k}_{\|}\right)}{\hbar\omega_c} = \varepsilon + \delta - \frac{\tilde{E}_{ex}\left(F_n,0\right)}{\hbar\omega_c} - \frac{n_c^2\hbar\omega_c}{M\left(F_n,B\right)c^2}\cdot\frac{x^2}{2}
\end{aligned} \tag{17}$$

The dispersion equation (16) in the case of circular polarization $\vec{\sigma}_{k_\uparrow}^-$ or $\vec{\sigma}_{k_\downarrow}^+$ in the denotations (17) looks as

$$\left(\varepsilon-\frac{x^2}{2}\right)=f_{osc}\left\{\frac{\left|a_0^-\right|^2\left|d_0^-\right|^2\left(1-\frac{x^2}{2}+\frac{7}{16}x^2\right)}{\varepsilon+\delta-\frac{\tilde{E}_{ex}\left(F_1,0\right)}{\hbar\omega_c}-\frac{n_c^2\hbar\omega_c}{M\left(F_1,B\right)c^2}\cdot\frac{x^2}{2}}+\right.$$

$$\frac{\frac{x^4}{16}}{\varepsilon+\delta-\frac{\tilde{E}_{ex}\left(F_4,0\right)}{\hbar\omega_c}-\frac{n_c^2\hbar\omega_c}{M\left(F_4,B\right)c^2}\cdot\frac{x^2}{2}}+ \tag{18}$$

$$\left.\frac{\left|b_1^-\right|^2\left(\frac{\pi l_0}{L_c}\right)^2\frac{x^6}{32}}{\varepsilon+\delta-\frac{\tilde{E}_{ex}\left(F_3,0\right)}{\hbar\omega_c}-\frac{n_c^2\hbar\omega_c}{M\left(F_3,B\right)c^2}\cdot\frac{x^2}{2}}+\frac{\left|a_0^-\right|^2\left|d_1^-\right|^2\left(\frac{\pi l_0}{L_c}\right)^2\frac{x^2}{2}\left(1-\frac{x^2}{2}+\frac{7}{16}x^4\right)}{\varepsilon+\delta-\frac{\tilde{E}_{ex}\left(F_5,0\right)}{\hbar\omega_c}-\frac{n_c^2\hbar\omega_c}{M\left(F_5,B\right)c^2}\cdot\frac{x^2}{2}}\right\}$$

For the opposite circular polarizations $\vec{\sigma}_{k_\uparrow}^{+}$ and $\vec{\sigma}_{k_\downarrow}^{-}$ we obtain

$$\left(\varepsilon-\frac{x^2}{2}\right)=f_{osc}\left\{\frac{\left(1-\frac{x^2}{2}+\frac{7}{16}x^4\right)}{\varepsilon+\delta-\frac{\tilde{E}_{ex}\left(F_4,0\right)}{\hbar\omega_c}-\frac{n_c^2\hbar\omega_c}{M\left(F_4,B\right)}\cdot\frac{x^2}{2}}+\frac{\frac{x^4}{16}\left|a_0^-\right|^2\left|d_0^-\right|^2}{\varepsilon+\delta-\frac{\tilde{E}_{ex}\left(F_1,0\right)}{\hbar\omega_c}-\frac{n_c^2\hbar\omega_c}{M\left(F_1,B\right)}\cdot\frac{x^2}{2}}+\right.$$

$$\left.+\frac{\left|b_1^-\right|^2\left(\frac{\pi l_0}{L_c}\right)^2\frac{x^2}{2}\left(1-\frac{x^2}{2}+\frac{7}{16}x^4\right)}{\varepsilon+\delta-\frac{\tilde{E}_{ex}\left(F_3,0\right)}{\hbar\omega_c}-\frac{n_c^2\hbar\omega_c}{M\left(F_3,B\right)c^2}\cdot\frac{x^2}{2}}+\frac{\left|a_0^-\right|^2\left|d_1^-\right|^2\left(\frac{\pi l_0}{L_c}\right)^2\frac{x^6}{32}}{\varepsilon+\delta-\frac{\tilde{E}_{ex}\left(F_5,0\right)}{\hbar\omega_c}-\frac{n_c^2\hbar\omega_c}{M\left(F_5,B\right)c^2}\cdot\frac{x^2}{2}}\right\} \tag{19}$$

The factors $\left(1-\frac{x^2}{2}+\frac{7}{16}x^4\right)$ and $x^4/16$ in both expressions (18) and (19) express the geometric selection rules (12). They differ from the values 1 and 0 correspondingly due to the non-zero in-plane component $\left|\vec{k}_{\|}\right|$ of the photon wave vector $\vec{k}=\pm\vec{a}_3\,(\pi/L_c)+\vec{k}_{\|}$. The origin of the factor $x^2/2$ is related with the quadrupole-active quantum transitions. The obtained energy spectra are discussed in the next section.

## 3 Energy spectrum of the magnetoexciton polaritons. Numerical calculations

To better understand the structure of the magnetoexciton-polariton branches, it is useful to represent the picture of the magnetoexciton energy levels alone. At the point $\vec{k}_{\|}$ they are shown in the Figure.1 for a wide range of the magnetic field strengths of up to 100 T for concrete values of the electron and whole g-factors, external electric field strength and the nonparabolicity parameter $C$. They were not demonstrated in the Refs.[2, 3]. The dispersion laws in Fig. 1 are represented for five magnetoexciton polariton branches and two magnetoexciton branches in the absence of the external electric field, $E_z = 0$. It means that the Landau quantization (LQ) takes place including the Zeeman splitting but without Rashba spin-orbit coupling and without nonparabolicity of the heavy-hole dispersion law.

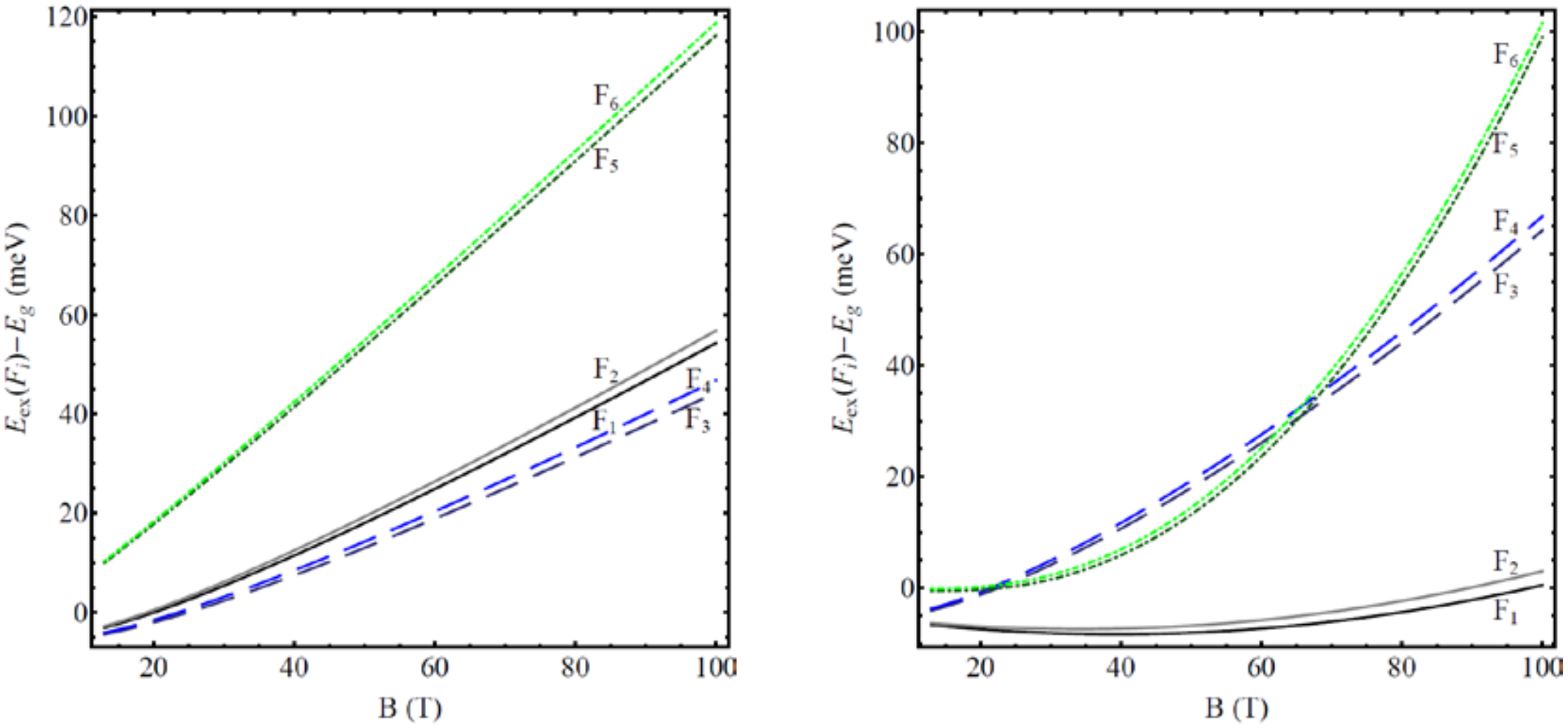

**Fig.1.** The energy levels of the six magnetoexciton states in the point $\vec{k}_{\|} = 0$ in dependence on the magnetic field strength $B$ at electron and hole g-factors $g_e = -0,5$ and $g_h = 2$ and at two different values of the external electric field strength $a) E_z = 0, C = 0$ and $b) E_z = 10\ \mathrm{kV/cm}, C = 20$.

Contrary to the previous calculations of Refs. [2, 3] the cavity photon resonance value $E_C$ was chosen the same at different values of the magnetic field strength $B$. The cavity detuning $\Delta$ of the value $E_c$ as regards the QW energy band gap $E_g^0$, which does not depend on the applied magnetic field was introduced $\Delta = E_c - E_g^0 = -2\text{mev}$. It means that the detuning between the cavity mode energy and the magnetoexciton energy $E_{ex}\left(F_n, 0\right)$ changes in dependence on the magnetic field strength. This variant is similar to that discussed in Ref. [17]. In contrast to the figures 5, 6, 7 of Ref. [2], where different combinations of the electron and hole g-factors were investigated, in the present variant we have chosen the electron and hole g-factors equal to the values $g_e = -0,5$ and $g_h = 2$, what leads to the exciton g-factor $g_{ex} = g_e + g_h = 1,5$. These values are very close to those determined in Ref. [17] in the case of $In_xGa_{1-x}As$ $QW$ with $x = 0,045$. The obtained polariton energy branches are presented in Figs. 2 and 3.

The dispersion laws are represented for four different values of the magnetic field strength $B$ changing in the interval from 10 T to 40T. In the interval 0-10T the Wannier-Mott structure of the exciton dominates, and the magnetoexcitons in GaAs-type $QW_s$ do not exist.

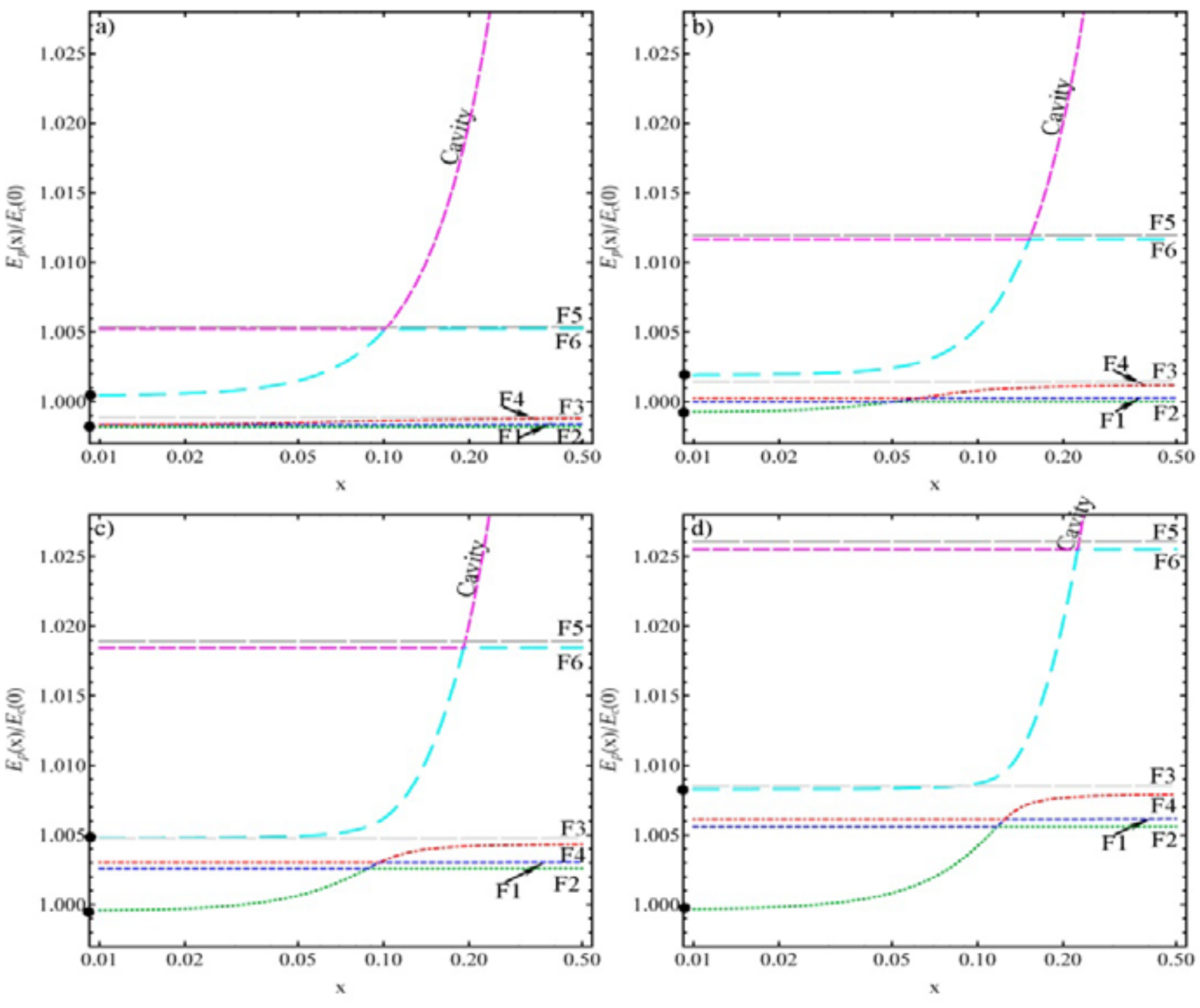

**Fig.2** The dispersion laws for five magnetoexciton-polariton branches and for two magnetoexciton branches forbidden in optical transitions for different values of the magnetic field strength as follows: $a)B = 10T;\ b)B = 20T;\ c)B = 30T;\ d)B = 40T$. The case of the absence of the external electric field $E_z = 0$ and $C = 0$.

The cavity resonance detuning as regards the QW energy band gap $E_g^0$ was chosen $\Delta = -2\text{mev}$.

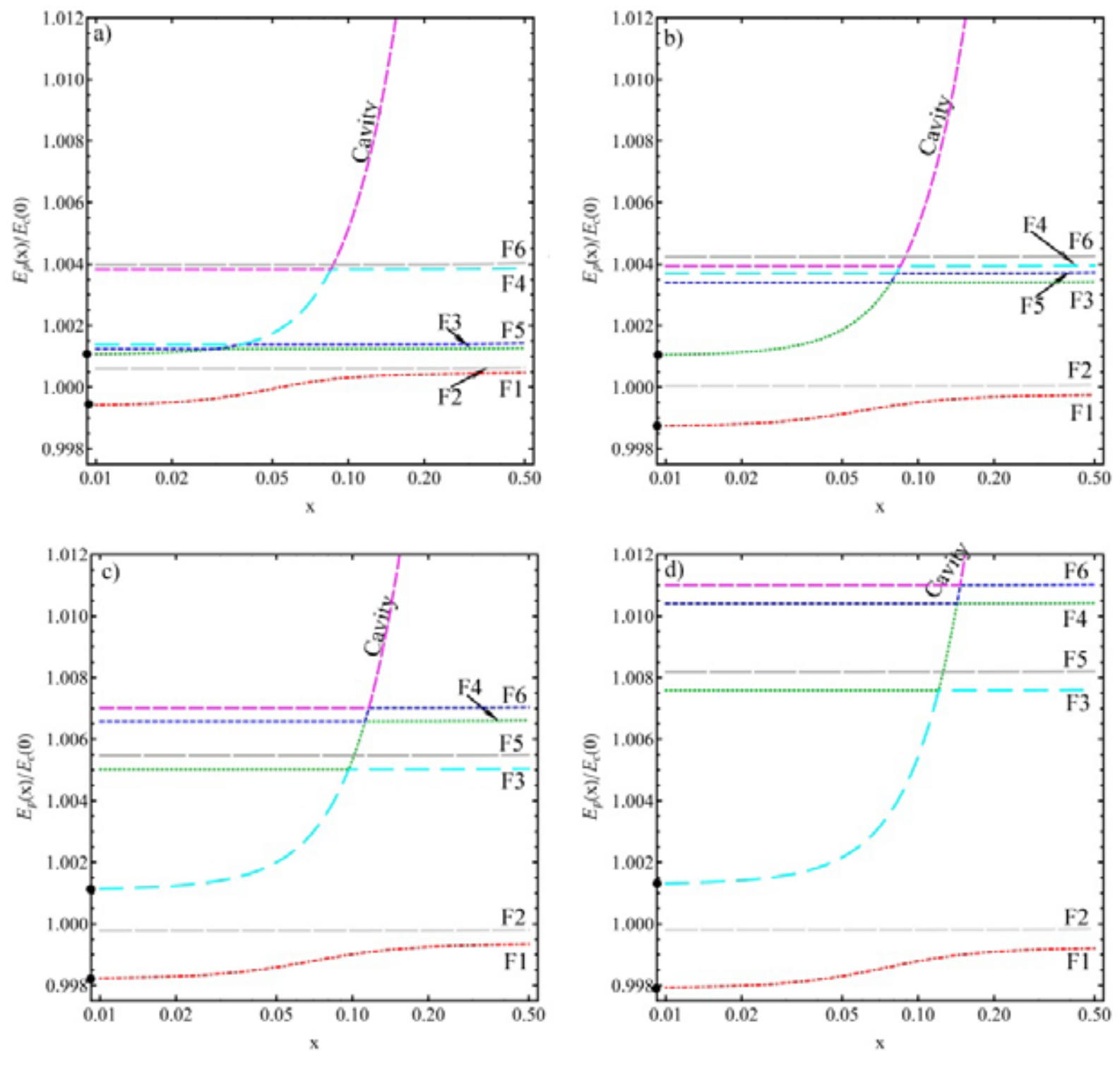

**Fig. 3** The dispersion laws for five magnetoexciton-polariton branches and for two magnetoexciton branches forbidden in optical transitions for different values of the magnetic field strength as follows: $a) B = 10T;\ b) B = 20T;\ c) B = 30T;\ d) B = 40T$. The case of the RSOC in the presence of the external electric field $E_z = 10\,\mathrm{kV/cm}$ and $C = 20$. The cavity resonance detuning as regards the QW energy band gap $E_g^0$ was chosen $\delta = -7\mathrm{mev}$. The points on the energy axes denote the positions of the upper and lower polariton branches in the point $x = 0$

The effective polariton mass for different polariton branches at the point $\left|\vec{k}_{\|}\right|$ can be calculated introducing the notations

$$
\begin{aligned}
&E_P\left(F_n,\vec{k}_{\|}\right) = E_c(0)\, f\left(F_n,x\right);\ x = \frac{\left|\vec{k}_{\|}\right| L_c}{\pi} \\
&m_{eff}\left(F_n,\vec{k}_{\|}\right) = \frac{\hbar^2}{\frac{d^2}{d\vec{k}_{\|}^2} E_P\left(F_n,\vec{k}_{\|}\right)} = m_c \frac{1}{f''\left(F_n,x\right)} \qquad (20) \\
&m_c = \frac{\hbar \pi n_c}{c L_c};\ E_c(0) = \hbar\omega_c = \frac{\hbar \pi c}{L_c n_c}
\end{aligned}
$$

Fig. 4 shows results of the numerical calculations for the lower polariton branches at the point $x = 0$.

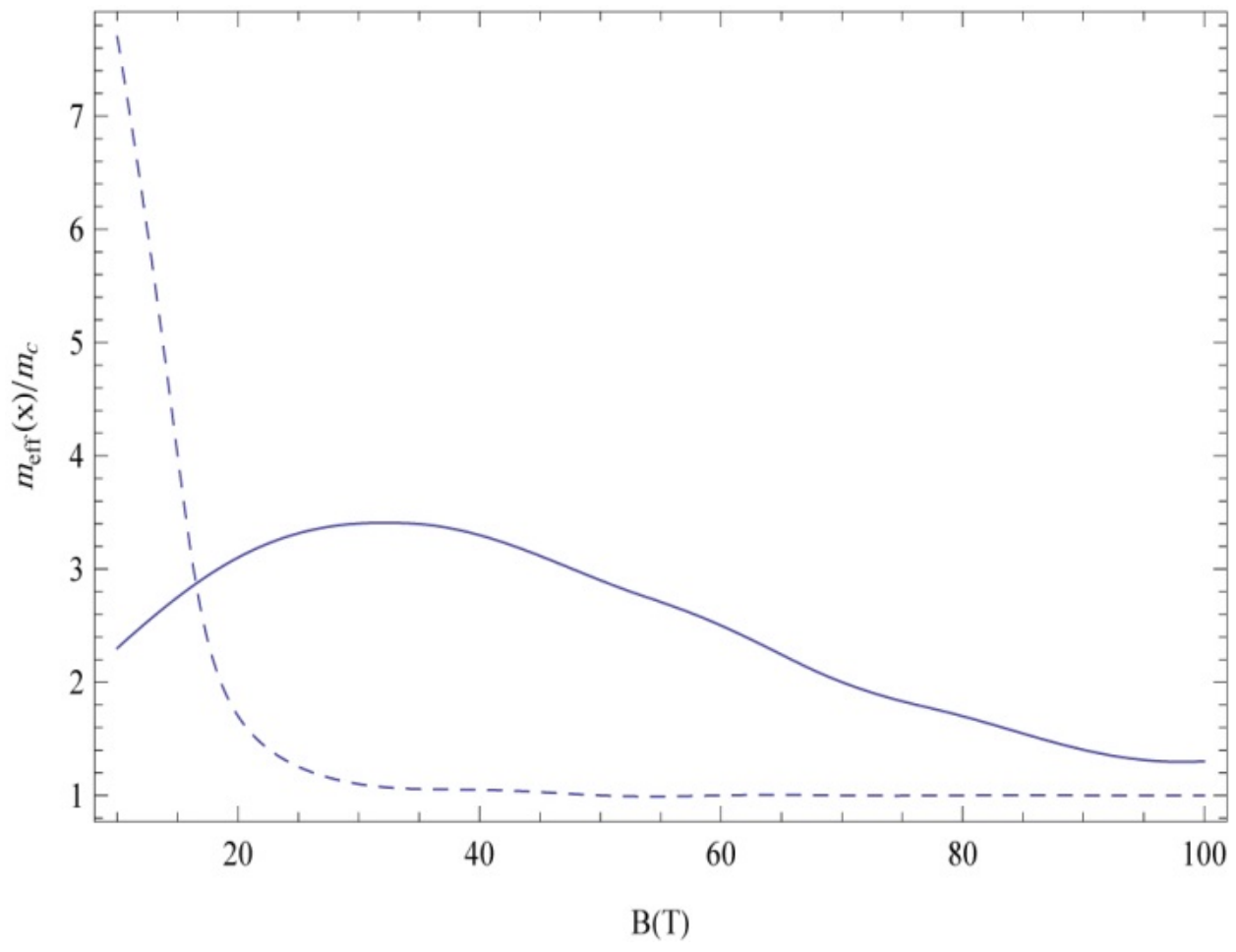

**Fig.4.** The dimensionless effective polariton mass $m_{eff}(0)/m_C$ in the point $x = 0$ of the lower polariton branches. In the case of the magnetoexciton state $F_1$ dipole-active in the quantum transition from the ground state of the crystal under the influence of the light with circular polarization $\vec{\sigma}^{-}_{\pi/L_c}$ in dependence on the magnetic field strength $B$. The Solid line corresponds to the presence of the RSOC with $E_z = 10\,\mathrm{kV/cm}$ and $C = 20$, whereas the dashed line in the absence of the RSOC ( $E_z = 0$, $C = 0$ )

The Rabi splitting of the two polariton energy branches is determined by the difference between the energies of the upper and lower polariton branches at the point $x = 0$. The splittings at different values of the magnetic field strength were determined using Figs. 2 and 3 and are shown in the Fig. 5.

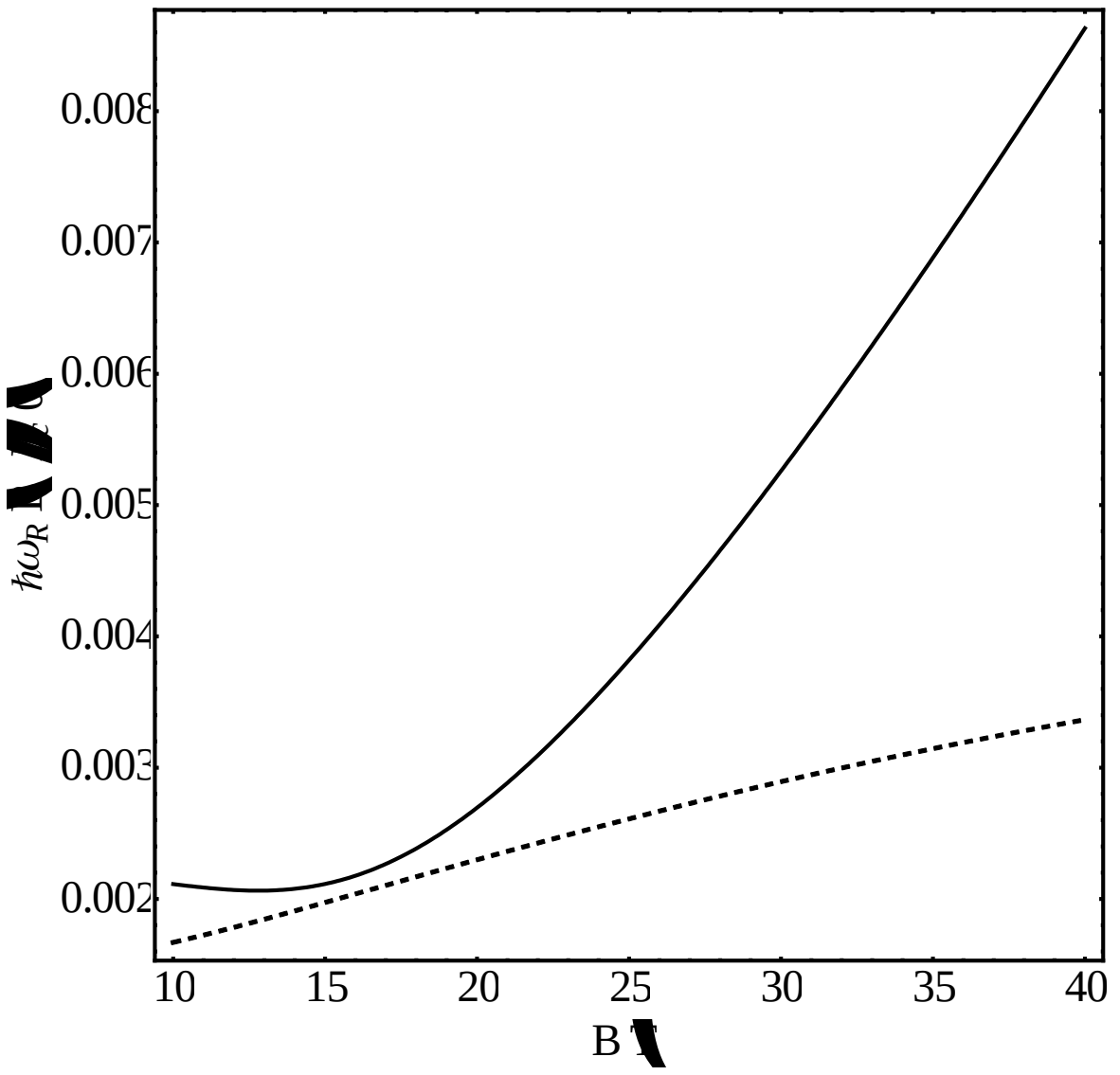

**Fig. 5.** The Rabi splittings between the upper and lower polariton branches in the point $x = 0$ at different values of the magnetic field strength. The solid line corresponds to the absence of the RSOC with $E_z = 0$ and $C = 0$. The dashed line corresponds to the presence of the RSOC with $E_z = 10\,\mathrm{kV/cm}$ and $C = 0$.

Fig. 6 shows the dispersion laws of the different magnetoexciton polariton energy branches in the range of magnetic field strength from 10 to 40 T for the case of RSOC with the parameters $E_z = 10\,\mathrm{kV/cm}$ and $C = 20$. In contrast to Figs. 1-4, here the electron g-factor $g_e = 1$ and the heavy-hole g-factor $g_h = 5$ were chosen in order to demonstrate the essential influence of the Zeeman splitting effects on the polariton branches structure.

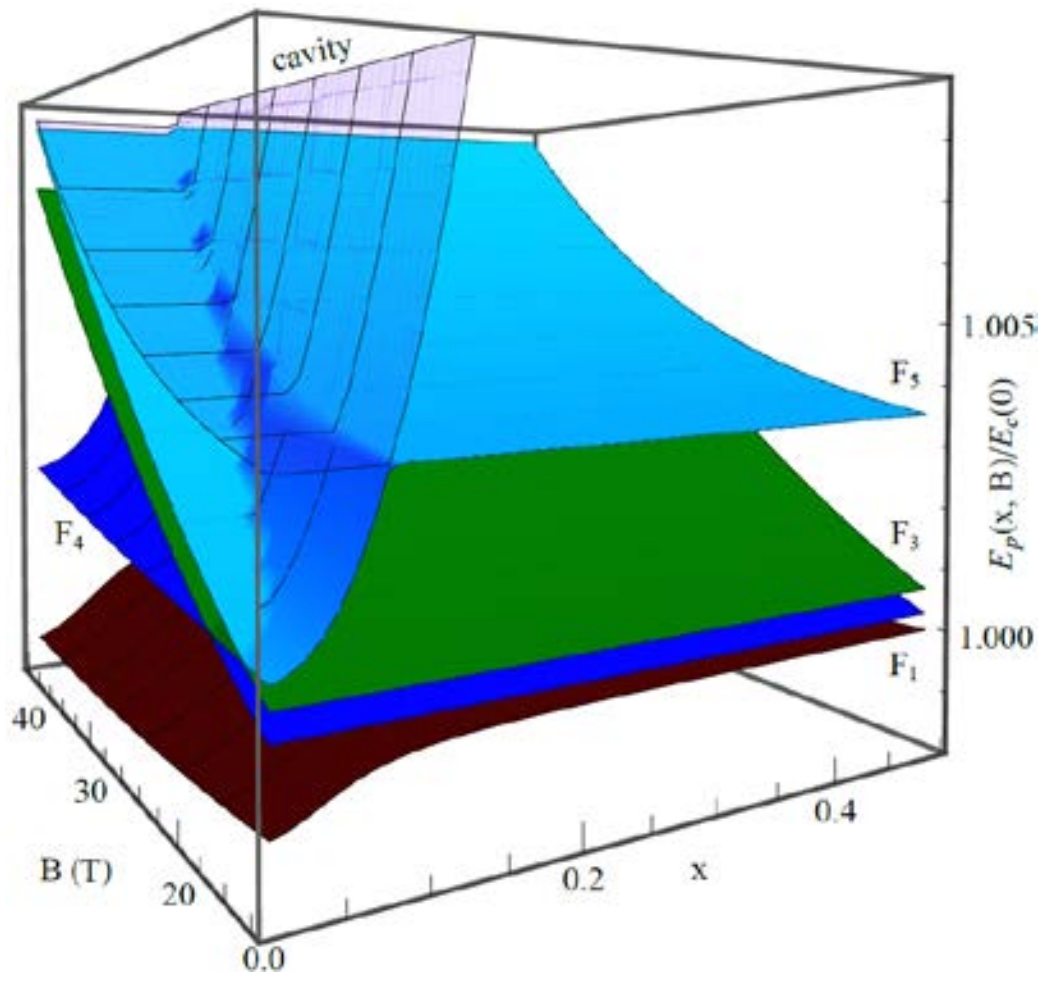

**Fig. 6.** The dispersion laws of the polariton energy branches at different values of the magnetic field strength taking into account the RSOC with parameters $E_z = 10\,\mathrm{kV/cm}$ and $C = 20$. The electron and heavy-hole g-factors with the values $g_e = 1$ and $g_h = 5$ were chosen.

At the point $x = 0$ the five order dispersion equations (18) and (19) become quadratic containing only one dipole-active magnetoexciton levels equal to $F_1$ and $F_4$, correspondingly, interacting with cavity photons with wave vector $\vec{k}_z = \pi/L_c$ and with two different circular polarizations $\vec{\sigma}^-_{\pi/L_c}$ and $\vec{\sigma}^+_{\pi/L_c}$, respectively. The energy values and the magnetic masses in these two cases are $E_{ex}(F_1,0), E_{ex}(F_4,0), M(F_1,B)$ and $M(F_4,B)$.

It should be mentioned that the equations (18) and (19) were obtained taking into account only the resonance terms of the exciton-photon interaction Hamiltonian and neglecting the antiresonance terms. This approximation is valid due to the cut-off the cavity-photon dispersion law and the absence of the photons with nearly zero energy, where the antiresonance terms could be important.

The dispersion equations (18) and (19) at the point $x = 0$ contain in their right hand sides only one term. Such simplified quadratic dispersion equations can be obtained using the quadratic Hamiltonians of the type:

$$\begin{aligned} H_2(0) &= E_{ex}(F,0)\hat{\psi}_{ex}^{+}(F,0)\hat{\psi}_{ex}(F,0) + \\ &+ E_c(0)C_{\pi/L_c,\vec{\sigma}}^{+} + \varphi\left(F,\vec{\sigma}_{\pi/L_c},0\right)\hat{\psi}_{ex}^{+}(F,0)C_{\pi/L_c,\vec{\sigma}} + \varphi^{*}\left(F,\vec{\sigma}_{\pi/L_c},0\right)C_{\pi/L_c,\vec{\sigma}}^{+}\psi_{ex}(F,0), \end{aligned} \qquad (21)$$

where $\left(F,\vec{\sigma}_{\pi/L_c},0\right)$ equals to $(F_1,\vec{\sigma}_{\pi/L_c}^{-})$ or to $(F_4,\vec{\sigma}_{\pi/L_c}^{+})$.

The alternative approach based on the Hamiltonian (21) is needed to introduce the polariton creation and annihilation operators and to deduce the corresponding Hopfield coefficients [26]. To this end we will start with the equations of motion for the magnetoexciton and cavity-photon operators in the stationary conditions

$$\begin{aligned} & i\hbar\frac{d}{dt}\psi_{ex}(F,0) = \left[\psi_{ex}(F,0),H_2(0)\right] = \hbar\omega\psi_{ex}(F,0) = \\ & = E_{ex}(F,0)\hat{\psi}_{ex}(F,0) + \varphi(F,0)C_{\pi/L_c,\vec{\sigma}} \\ & i\hbar\frac{d}{dt}C_{\pi/L_c,\vec{\sigma}} = \left[C_{\pi/L_c,\vec{\sigma}},H_2(0)\right] = \hbar\omega C_{\pi/L_c,\vec{\sigma}} = \\ & = E_C(0)C_{\pi/L_c,\vec{\sigma}} + \varphi^{*}(F,0)\hat{\psi}_{ex}(F,0) \end{aligned} \qquad (22)$$

They lead to the algebraic relations between the operators, to the second order dispersion equation and to the upper and lower polariton branches as follows

$$\begin{aligned} & \left(\hbar\omega - E_{ex}(F,0)\right)\hat{\psi}_{ex}(F,0) = \varphi(F,0)C_{\pi/L_c,\vec{\sigma}} \\ & \left(\hbar\omega - E_C(0)\right)C_{\pi/L_c,\vec{\sigma}} = \varphi^{*}(F,0)\hat{\psi}_{ex}(F,0) \\ & \left(\hbar\omega - E_{ex}(F,0)\right)\left(\hbar\omega - E_C(0)\right) = \left|\varphi(F,0)\right|^2 \\ & \hbar\omega_{\substack{U \\ L}} = \frac{E_{ex}(F,0)+E_C(0)}{2} \pm \sqrt{\left(E_{ex}(F,0)-E_C(0)\right)^2 + 4\left|\varphi(F,0)\right|^2} \end{aligned} \qquad (23)$$

Now we introduce the polariton creation and annihilation operators corresponding to the upper and lower polariton branches at the point $x = 0$, using the Hopfield coefficients [26].

$$\hat{P}_u = u\hat{\psi}_{ex}\left(F,0\right) + vC_{\frac{\pi}{L_c},\vec{\sigma}}; \hat{P}_L = \xi C_{\frac{\pi}{L_c},\vec{\sigma}} + \eta\hat{\psi}_{ex}\left(F,0\right)$$

$$\left[\hat{P}_u, \hat{P}_v^+\right] = 1, \ \left[\hat{P}_L, \hat{P}_L^+\right] = 1, \ \left[P_L, P_u^+\right] = \left[P_u, P_L^+\right] = 0 \quad (24)$$

$$|u|^2 + |v|^2 = 1; \ |\xi|^2 + |\eta|^2 = 1$$

The equations of motion for the operators $\hat{P}_u$ and $\hat{P}_L$ and the algebraic relations between the Hopfield coefficient are

$$i\hbar\frac{d\hat{P}_u}{dt} = \hbar\omega_u\hat{P}_u; \ i\hbar\frac{d\hat{P}_L}{dt} = \hbar\omega_L\hat{P}_L$$

$$u = \frac{\varphi^* v}{\left(\hbar\omega_u - E_{ex}\left(F,0\right)\right)}; \ \mathrm{v} = \frac{\varphi u}{\left(\hbar\omega_u - E_C\left(0\right)\right)} \quad (25)$$

$$\xi = \frac{\varphi\eta}{\left(\hbar\omega_L - E_C\left(0\right)\right)}; \ \eta = \frac{\varphi^*\xi}{\left(\hbar\omega_L - E_{ex}\left(F,0\right)\right)}$$

Taking into account the equalities $\left(\hbar\omega_u - E_x\left(F,0\right)\right)^2 = \left(\hbar\omega_L - E_C\left(0\right)\right)^2$ and $\left(\hbar\omega_u - E_C\left(0\right)\right)^2 = \left(\hbar\omega_L - E_{ex}\left(F,0\right)\right)^2$ we can write

$$|u|^2 = |\xi|^2 = \frac{|\varphi(F,0)|^2}{\left(\hbar\omega_u - E_{ex}\left(F,0\right)\right)^2 + |\varphi(F,0)|^2} =$$

$$= \frac{1}{2}\left[1 + \frac{E_{ex}\left(F,0\right) - E_C\left(0\right)}{\sqrt{\left(E_{ex}\left(F,0\right) - E_C\left(0\right)\right)^2 + 4|\varphi(F,0)|^2}}\right],$$

$$|v|^2 = |\eta|^2 = \frac{|\varphi(F,0)|^2}{\left(\hbar\omega_u - E_C\left(0\right)\right)^2 + |\varphi(F,0)|^2} = \quad (26)$$

$$= \frac{1}{2}\left[1 - \frac{E_{ex}\left(F,0\right) - E_C\left(0\right)}{\sqrt{\left(E_{ex}\left(F,0\right) - E_C\left(0\right)\right)^2 + 4|\varphi(E,0)|^2}}\right]$$

The coefficient $|u|^2$ and $|v|^2$ calculated for the magnetoexciton state $F_1$ and cavity-photons with circular polarization $\vec{\sigma}^-_{\pi/L_c}$ are

$$\begin{vmatrix} |u|^2 \\ |v|^2 \end{vmatrix} = \frac{1}{2}\left[ 1 \pm \frac{\left( \frac{\tilde{E}_{ex}(F_1,0)}{E_C(0)} - \delta \right)}{\sqrt{\left( \frac{\tilde{E}_{ex}(F_1,0)}{E_C(0)} - \delta \right)^2 + 4\, fosc \left| c_0^- \right|^2 \left| d_0^- \right|^2}} \right] \tag{27}$$

Fig. 7 shows the Hopfield coefficients at the point $x = 0$ determining the composition of the polariton creation and annihilation operators. They were calculated in the rotation wave approximation, when the antiresonance terms in the magnetoexciton-photon interaction Hamiltonian can be neglected.

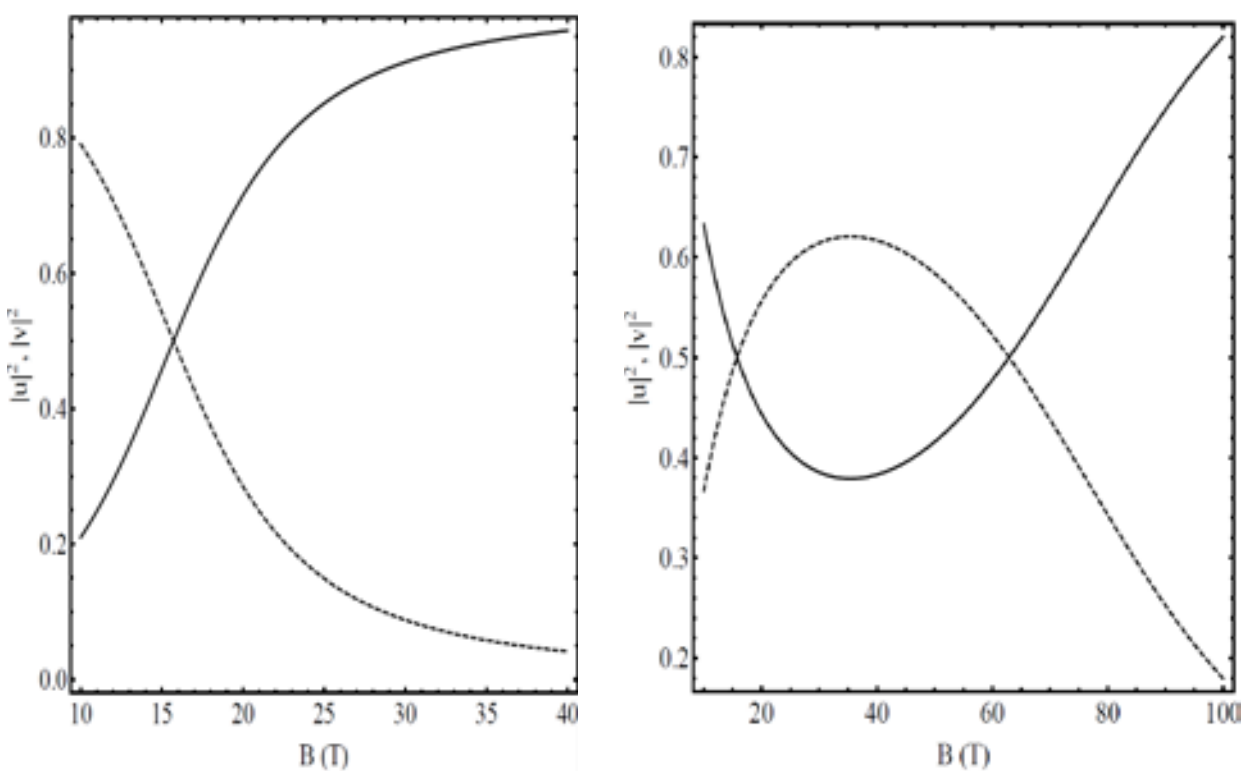

**Fig. 7** The dependences on the magnetic field strength $B$ of the Hopfield coefficients square moduli $|u|^2$ and $|v|^2$ in the case of magnetoexciton state $F_1$ interacting with the cavity photons with wave vector $k_z = \pi/L_C$, $x = 0$ and circular polarization $\vec{\sigma}^-_{\pi/L_c}$ in two cases: a) in the absence of the RSOC $(E_z = 0, C = 0)$; b) in the presence of the RSOC with $E_z = 10\,kV/cm$ and $C = 20$. The coefficients $|u|^2 = |\xi|^2$ are represented by the solid lines, whereas the coefficients $|v|^2 = |\eta|^2$ by the dashed lines.

Let us compare the results shown in figs. 7 and 4, first - in the absence of the RSOC and the cavity detuning $\Delta = -2\text{mev}$. One can observe that in the range of magnetic field strength from 10 to 20 T the photon component of the LPB in the point $x = 0$ $|u(0)|^2$ increases from the value 0,2 to 0,8. It leads to reduction of the effective mass of the LPB from 7 to 1 $m_c$ . In the presence of the RSOC and of the cavity detuning $\Delta = -7\text{mev}$ the value $|u(0)|^2$ depends nonmonotonically on the magnetic field strength with a minimal value 0,4 for $B \approx 30\text{T}$ . In this point the effective mass on the LPB has a maximum value 3,5 $m_c$ . With the further increase of the magnetic field the effective mass decreases up to the limiting value $m_c$ .

The nonmonotonic dependence of the Hopfield coefficients square moduli $|u|^2 = |\xi|^2$ and $|v|^2 = |\eta|^2$ from the magnetic field strength are directly related with the similar dependence of the magnetoexciton energy levels shown in Fig. 3b, as well as with the selection of the cavity detuning value $\Delta = -7mev$ . In this case the energy of the cavity mode is situated on the energy scale in the surrounding of the polariton branch $F_1$ . As a result the numerator of the fraction in the square bracket of the formula (27) vanishes at two different values of the magnetic field. At these points the square bracket at (27) equals to 1 and the coefficients $|u|^2 = |v|^2 = 1/2$ .

The magnetoexciton state $F_1$ has the electron in the spinor state $(e, R_1)$ , where the spin projection $S_z^e$ equals to ½ with the probability $|a_0^-|^2$ , and equals to -½ with the probability $|b_1^-|^2$ . The heavy-hole state $(h, \varepsilon_3^-)$ is characterized by the values $\tilde{S}_z^h = -1/2$, $\text{j}_z^h = -3/2$ and $M_h = -1$ with the probability $|d_0^-|^2$ and by the values $\tilde{S}_z^h = 1/2$, $\text{j}_z^h = 3/2$ and $M_h = 1$ with the probability $|c_3^-|^2$ . The sum of the exciton angular momentum projection $\Sigma_z^{ex} = S_z^e + j_z^h$ equals to the value $-1$ with the probability $|a_0^-|^2 |d_0^-|^2$ . The orbital selection rule for the quantum numbers of the Landau quantization levels in the case of the dipole-active transitions is $n_e = n_h$. It is held for the case $\left|a_0^-\right|^2 \gg \left|b_1^-\right|^2$ and $\left|d_0^-\right|^2 \gg \left|c_3^-\right|^2$.

The magnetoexciton state $F_2 = \left(e, R_2; h_1 \varepsilon_3^-\right)$ has the electron spinor wave function with $S_z^e$ projection $S_z^e = -1/2$, whereas the heavy-hole spinor wave function is the same as in the case of the magnetoexciton state $F_1$. In this case we have the sum of the exciton angular momentum projection $\Sigma_z^{ex} = -2$ with probability $\left|d_0^-\right|^2$ and $\Sigma_z^{ex} = 1$ with probability $\left|c_3^-\right|^2$. The first case corresponds to the dark WME state and is forbidden due to the spin selection rules.

In the second case difference in the numbers of the Landau quantization energy levels $n_e = 0$ and $n_h = 3$ is too large and the corresponding quantum transition is forbidden in the both dipole and quadrupole approximations.

The magnetoexciton state $F_3 = \left(e, R_i; h, \varepsilon_0\right)$ differs from the state $F_1$, by the heavy-hole spinor state $\left(h, \varepsilon_0\right)$ with $\tilde{S}_z^h = 1/2$, $j_z^h = 3/2$ and $M_h = 1$.

The exciton angular momentum projection $\Sigma_z^{ex} = 2$ has probability $\left|a_0^-\right|$ and $\Sigma_z^{ex} = 1$ has probability $\left|b_1^-\right|^2$. This quantum transition is forbidden exclusively due to spin selection rules in the first case and it is allowed in the second case in the quadrupole approximation, because the numbers $n_e = 1$ and $n_h = 0$ differ by one. The magnetoexciton state $F_4 = \left(e, R_2; h, \varepsilon_0\right)$ is characterized by the quantum numbers $S_z^e = -1/2$, $\tilde{S}_z^h = 1/2$, $j_z^h = 3/2$, $M_h = 1$ and $\Sigma_z^{ex} = 1$. It is dipole-active, satisfying to the orbital selection rule $n_e = n_h = 0$ and is the heir of the bright WME state with $\Sigma_z^{ex} = 1$.

The magnetoexciton state $F_5 = \left(e, R_1, h, \varepsilon_4^-\right)$ differs from the state $F_1$ by the heavy-hole state $\left(h, \varepsilon_4^-\right)$ instead of $\left(h, \varepsilon_3^-\right)$. The exciton angular momentum projection $\Sigma_z^{ex} = -1$ gas the probability $\left|a_0^-\right|^2 \left|d_1^-\right|^2$. This quantum transition is quadrupole active due to the difference by one between the quantum numbers of the Landau quantization levels $n_e = 0$ and $n_h = 1$.

Unlike the state $F_5$ the magnetoexciton state $F_6 = \left(eR_2; h, \varepsilon_4^-\right)$ has the electron state with $S_z^e = -1/2$, what leads to the value $\Sigma_z^{ex} = 1$ with the probability $\left|c_4^-\right|^2$. But due to a great difference between the numbers $n_e = 0$ and $n_h = 4$ of the corresponding Landau quantization levels the optical quantum transition from the ground state of the crystal to the magnetoexciton state $F_6$ is forbidden in the both dipole and quadrupole approximations.

The magnetoexciton states $F_1$ and $F_4$ have the total angular momentum projections $\Sigma_z^{ex}$ equal to -1 and +1, correspondingly. They are the heirs the bright 2D Wannier-Mott exciton states with the same values of $\Sigma_z^{ex}$. In the case of 2D $WME_s$ these two states with the same values are splitted due to the electron-hole long-range Coulomb interaction. One can remember that in the case of the dipole-active triple-degenerate 3D exciton states in the cubic crystals the long-range Coulomb interaction gives rise to the transverse-longitudinal splitting of these exciton states. In the case of 2D magnetoexciton states $F_1$ and $F_4$, they are splitted even in the absence of the long-range Coulomb interaction, which was not taken into account in their construction. Fig. 8 shows splittings between the branches $F_1$ and $F_4$ at the point $x = 0$.

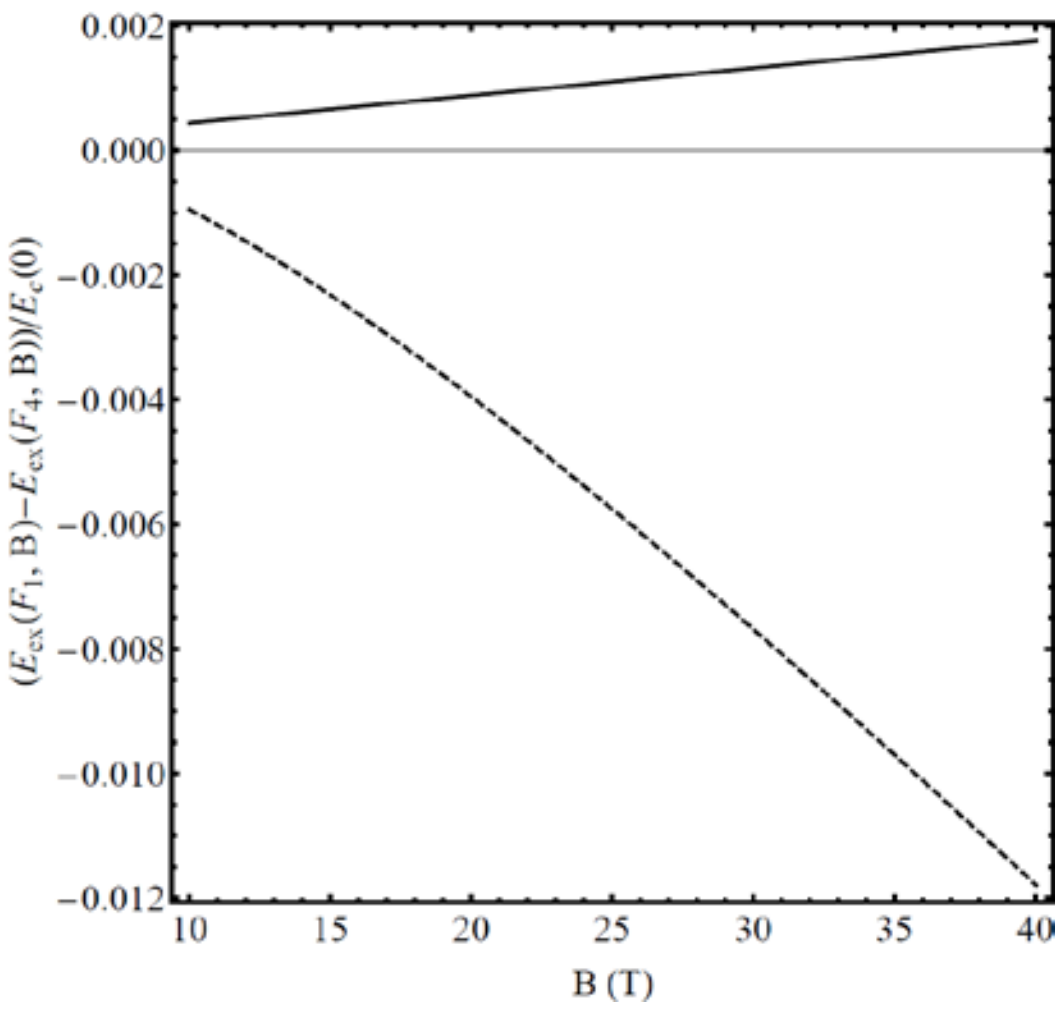

**Fig. 8.** The splittings between the two dipole-active magnetoexciton energy branches $F_1$ and $F_4$ in the point $x = 0$. The splitting is represented by the solid line in the absence of the RSOC, and by the dashed line in the case of its presence with the parameters $E_z = 10\,\mathrm{kV/cm}$ and $C = 20$.

The two splitted cavity polariton states, which determine the polarization of the emitted light in the case of 2D WME, were described using the pseudospin model. It can be considered as a basic element for the spinoptronics [27, 28]. The same description can be used in the case of magnetoexciton polaritons.

## 4 Conclusions

The dispersion laws of the two-dimensional cavity magnetoexciton-polaritons were described. The energy of the cavity resonance was tuned to the semiconductor band gap edge, whereas the detuning between the cavity photons and the magnetoexciton energy branches changed with the magnetic field strength. The Rabi splitting between the upper and the lower polariton branches at the point with in-plane wave vector $\vec{k}_{\parallel} = 0$ was determined in both cases corresponding to the presence and to the absence of the RSOC. In the first case the strength $E_z = 10\,\mathrm{kV/cm}$ of the external electric field perpendicular to the quantum well plane and the parameter of the nonparabolicity $C = 20$ of the heavy-hole dispersion law were chosen. The effective polariton mass at the point $\vec{k}_{\|} = 0$ of the lower polariton branch was determined in dependence on the magnetic field strength in both cases concerning the RSOC mentioned above.

In the same conditions the square moduli of the Hopfield coefficients, which determine the magnetoexciton and cavity photon components of the polariton branches in the point $\vec{k}_{\|} = 0$, were calculated. They have a nonmonotonous dependence on the magnetic field strength in the presence of the RSOC. It is the consequence of the nonmonotonous dependence on the magnetic field strength of the magnetoexciton energy branches when the cavity mode energy is lying on the energy scale in the surrounding of the selected magnetoexciton branch.

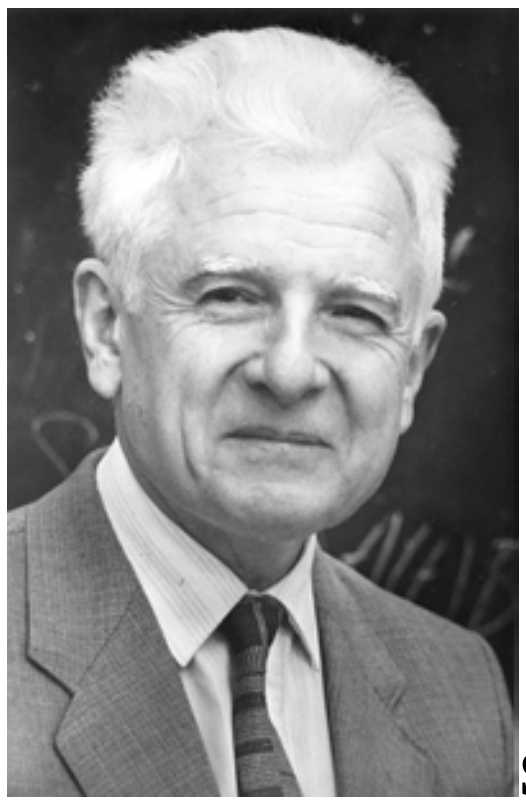

**Sveatoslav A.Moskalenko** was born on September 26, 1928 in the village Bravicha Calarash district, Republic of Moldova. Citizenship of Moldova. Moskalenko S.A. has graduated the Kishinev State University in 1951. In 1956 - 1959 he was post-graduate student of the Institute of Physics of the Academy of Sciences of Ukrainian SSR. He became Candidate of Physico-mathematical Sciences equivalent to PHD degree in 1961. Doctor of Physico-mathematical Sciences and the professor in the field of theoretical and mathematical physics beginning with 1971 and 1974 correspondingly. He became Laureate of the State prize in the field of Sciences and Technics of MSSR and USSR in 1981 and 1988 correspondingly. He was elected member-correspondent and full member of the Academy of Sciences of Moldova in 1989 and 1992 correspondingly.

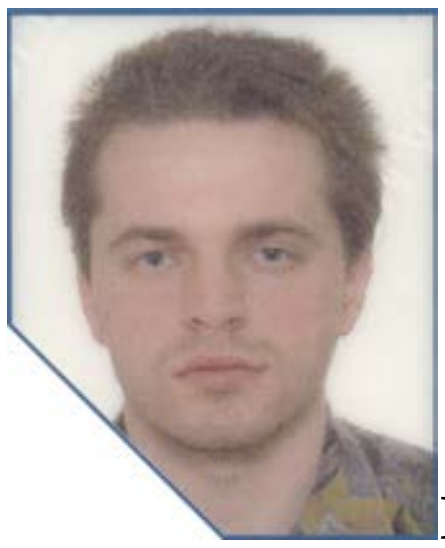

**Igor V. Podlesny** has graduated from the College of Informatics of Moldova (Mathematics and Informatics) in 2000; from the State University of Moldova (Theoretical Physics) in 2004. In 2004-2007 he was the postgraduate student in the Institute of Applied Physics of the Academy of Sciences of Moldova. Professional activity: 2003-Present, Institute of Applied Physics, Research worker. Field of professional activity: Condensed matter physics (magnetoexcitons), Theoretical physics. Participation in scientific international forums: 2004-Present, more than 30 conferences. He has participated in more than 5 national and international projects.

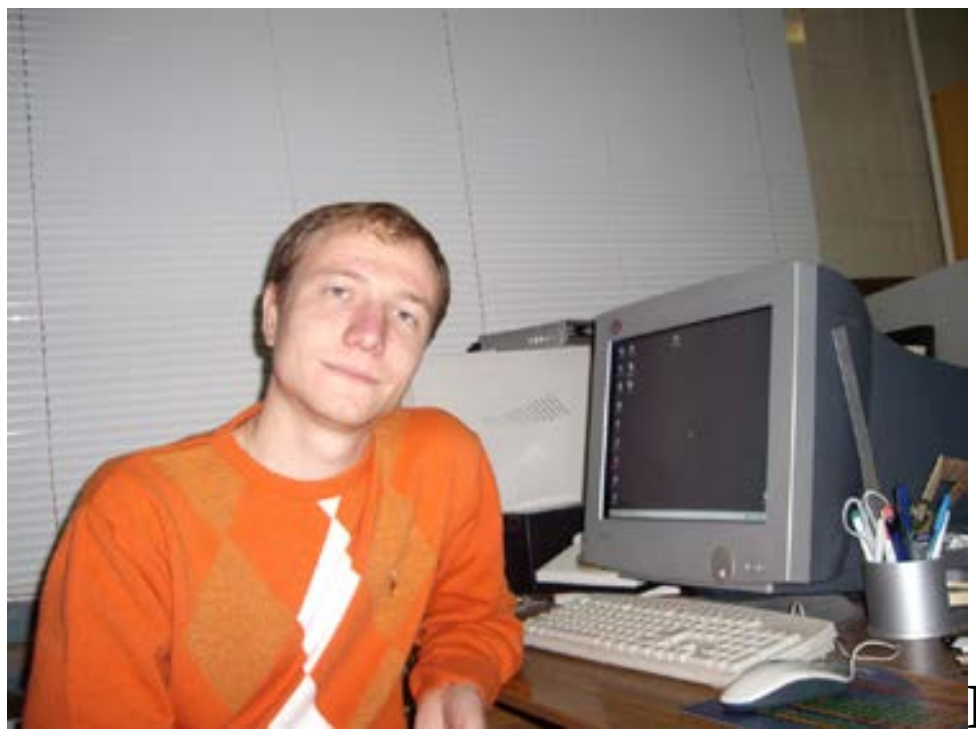

**Evgheni V. Dumanov** was born on September 30, 1982 in the Chisinau city, Republic of Moldova. E.V.Dumanov has graduated the Moldova State University in 2004. In 2004 – 2007 he was post-graduate student of the Institute of Applied Physics of the Academy of Sciences of Moldova. He received his Ph.D. in 2008 from Institute of Applied Physics in Chisinau for the theoretical and mathematical physics. In the Institute of Applied Physics of ASM, he is Senior Associate collaborator now. His research interests are collective properties of elementary particle in semiconductors, effects of Bose-Einstein condensation 2D magnetoexcitons, collective elementary excitation.

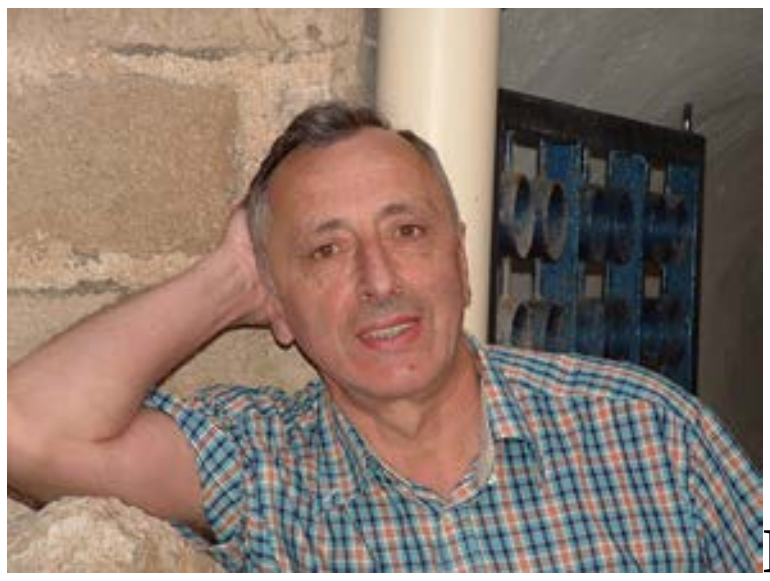

**Michael A. Liberman** was born in Moscow, USSR on October 23, 1942. He graduated from Moscow State University in 1966. From 1969 to 2003 he worked at P. Kapitsa Institute for Physical Problems, Academy of Sciences USSR. He received his Ph.D. in 1971 from P. Lebedev Physical Institute in Moscow for the group theory in quantum mechanics and invariant expansion of the relativistic amplitudes, and then his Doctor of Physical and Mathematical Sciences degree in 1981 for a thesis on ionizing shock waves. Since 1991 till 2009 he was professor of theoretical statistical physics at the Physics Department, Uppsala University, Sweden and presently he is a professor at the Nordic Institute for Theoretical Physics (NORDITA) KTH and Stockholm University. He is a citizen of both Russia and the Sweden.

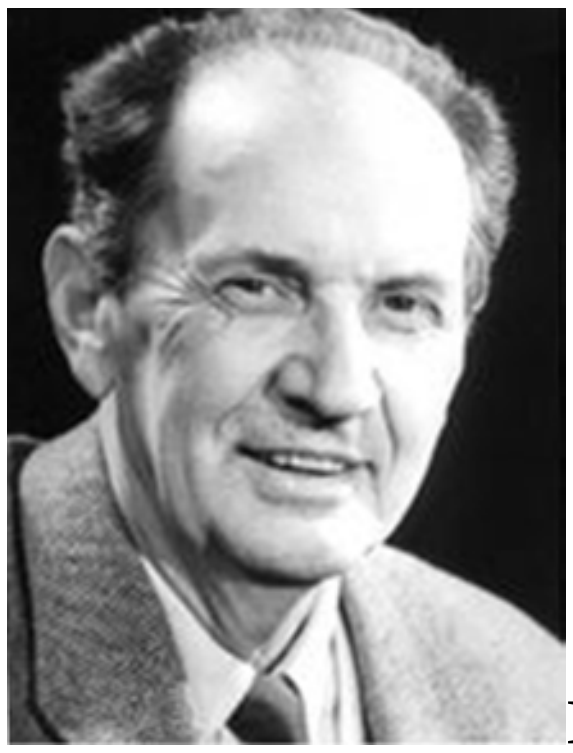

**Boris V. Novikov** graduated in 1956 year the Sanct-Petersburg (Leningrad) state university. From 1986 up till 2010 B.V. Novikov headed the chair of the Solid State Physics. In 1987 year the branch of this chair was organized in the frame of the A.F. Ioffe Physico-Technical Institute of the Russian Academy of Sciences (RAS). In 2005 year B.V. Novikov initiated the creation of the Russian-Germany Scientific Center. Boris V. Novikov is outstanding specialist in the optics of semiconductors and nanostructures. He is the author of more than 200 papers, inclusive of 2 monographs. He was a supervisor of 26 candidatus scientiarum theses and 4 his disciples became doctor of Sciences. B.V. Novikov was awarded by the premium of A.F. Ioffe RAS.

**Caption List**

**Fig. 1** The energy levels of the six magnetoexciton states.

**Fig. 2** The dispersion laws for five magnetoexciton-polariton branches and for two magnetoexciton branches forbidden in optical transitions for different values of the magnetic field strength.

**Fig. 3** The dispersion laws for five magnetoexciton-polariton branches and for two magnetoexciton branches forbidden in optical transitions for different values of the magnetic field strength.

**Fig. 4** The dimensionless effective polariton mass $m_{eff}(0)/m_C$ in the point $x = 0$ of the lower polariton branches

**Fig. 5** The Rabi splittings between the upper and lower polariton branches in the point $x = 0$ at different values of the magnetic field strength

**Fig. 6** The dispersion laws of the polariton energy branches at different values of the magnetic field strength taking into account the RSOC with parameters $E_z = 10\,\mathrm{kV/cm}$ and $C = 20$.

**Fig. 7** The dependences on the magnetic field strength $B$ of the Hopfield coefficients square moduli $|u|^2$ and $|v|^2$

**Fig. 8** The splittings between the two dipole-active magnetoexciton energy branches $F_1$ and $F_4$ in the point $x = 0$.